\documentstyle[12pt,preprint,aps]{revtex}

\begin{document}

\title{Gravity-driven instability in a spherical Hele-Shaw cell}

\author{Jos\'e A. Miranda, Fernando Parisio, Fernando Moraes}
\address{Laborat\'{o}rio de F\'{\i}sica Te\'{o}rica e Computacional,
Departamento de F\'{\i}sica,\\ Universidade Federal de Pernambuco,
Recife, PE  50670-901 Brazil}

\author{Michael Widom}
\address{Department of Physics, Carnegie Mellon University, 
Pittsburgh, PA  15213 USA}
\date{\today}
\maketitle

\begin{abstract}
A pair of concentric spheres separated by a small gap form a spherical
Hele-Shaw cell.  In this cell an interfacial instability arises when
two immiscible fluids flow. We derive the equation of motion for the
interface perturbation amplitudes, including both pressure and gravity
drivings, using a mode coupling approach.  Linear stability analysis
shows that mode growth rates depend upon interface perimeter and
gravitational force.  Mode coupling analysis reveals the formation of
fingering structures presenting a tendency toward finger tip-sharpening.

\end{abstract}

\section{Introduction}
\label{intro}

The Saffman-Taylor instability~\cite{Saf} has been the object of
extensive study during the last four decades~\cite{Rev}.  It arises at
the interface separating two viscous fluids constrained to flow in the
narrow gap between closed spaced parallel plates, a device known as
Hele-Shaw cell.  The cell thickness is smaller than any other length
scale in the problem, so the flow is effectively two-dimensional.  The
instability arises either from a pressure gradient advancing the less
viscous fluid against the more viscous one, or by gravity acting on the
density difference between the fluids.  The action of such
driving-force mechanisms leads to the celebrated viscous fingering
patterns~\cite{Saf,Rev}.

Most Saffman-Taylor investigations analyze flow between 
{\it flat} Hele-Shaw cells. In a previous work~\cite{Parisio} we 
started studying the Saffman-Taylor problem on 
curved surfaces by considering flow in a {\it spherical} Hele-Shaw 
cell (figure 1). The interfacial instability was produced 
by a nonzero flow injection rate $Q$, and gravitational effects 
were completely negleted. We examined the effect of cell 
curvature on the shape of the patterns and showed that 
positive spatial curvature inhibits finger tip-splitting.

In the present paper we focus on the influence of gravity in a
spherical Hele-Shaw cell. We consider both $Q=0$ and $Q>0$ at fixed
cell curvature.  The unperturbed domain shape is a polar cap of some
size, presenting an initially nearly circular boundary, which is
gravitationally unstable and evolves at constant area ($Q=0$), or
slowly increasing area ($Q>0$), without change of topology.

The study of viscous flow in a nonplanar Hele-Shaw cell is of interest
for both scientific and practical reasons.  On the scientific level,
the influence of spatial curvature on hydrodynamic flow is a matter of
fundamental interest.  It also provides a simple mathematical model to
describe more general situations involving the filling of a thin
cavity between two walls of a given shape with fluid. On the practical
level, it may have applications in a number of industrial,
manufacturing processes, ranging through pressure moulding of molten
metals and polymer materials~\cite{Rich}, and formation of coating
defects in drying paint thin films~\cite{Schwartz}.
 
The gravity-driven instability on a spherical Hele-Shaw cell also
allows one to gain insight into the properties of the dynamically
similar, but more complex, Rayleigh-Taylor
instability~\cite{Ray,Tay,Chan} for flow on substrates of arbitrary
shapes. One familiar example of this type of instability is the
formation of fingering patterns when chocolate syrup drains, due to
the action of gravity, from the top of a scoop of ice cream.  Despite
the apparent simplicity of this example, it is in fact a rather
complicated three-dimensional problem~\cite{Schwartz}, much less
amenable to analytic treatment than its two-dimensional Hele-Shaw
counterpart.

The geometrically constrained spherical Hele-Shaw cell forces the flow
to become essentially two-dimensional, and the interface
one-dimensional. High viscosity eliminates inertial terms from the
equations of motion, making the problem simpler yet. In contrast, the
conventional Rayleigh-Taylor problem is inertially driven and
three-dimensional effects become important~\cite{Ray,Tay,Chan}.

The outline of the work is the following: section~\ref{derivation}
derives the nonlinear equation of motion including both injection and
gravitational driving. Section~\ref{discussion} discusses the
resulting motion.  Section~\ref{linear} considers linear stability
analysis for purely gravitational driving with $Q=0$. The growth is
purely exponential, and the linear growth rate depends on the geodesic 
distance from the sphere's north pole. Larger distance causes 
both faster growth and more unstable modes, though the 
characteristic wavelength is nearly constant. The
nonvanishing injection case $Q>0$ is studied in section~\ref{inject}.
Injection of a high viscosity fluid tends to stabilize the
interface. For $Q>0$ linear growth is non-exponential due to evolution
of the linear growth rate with distance. Section~\ref{nonlinear}
studies the coupling of a small number of modes. It is shown that for
gravity-driven flow fingers have tendency to finger tip-sharpening.

\section{Stability analysis and mode coupling} 
\label{derivation}


Consider two immiscible, incompressible, viscous fluids, flowing in a 
spherical Hele-Shaw cell of thickness $b$ (see figure 1). 
The effetively two-dimensional flow takes place on the surface of 
a sphere endowed with the metric~\cite{Dub}
\begin{equation}
\label{metric}
ds^{2}=d \rho^{2} + a^{2} \sin^{2} \left (\frac{\rho}{a} \right ) d\varphi^{2},
\end{equation}
where $a$ is the radius of curvature of the sphere, $0 \le \varphi < 2
\pi$ denotes the azimuthal angle measured on the sphere and $0 \le \rho
\le \pi a$ is the geodesic distance from the sphere's north pole. The polar angle $\theta=\rho/a$.
Denote the viscosities and densities of the upper and lower fluids,
respectively as $\eta_{1}$, $\varrho_{1}$ and $\eta_{2}$,
$\varrho_{2}$. Consider the case $\varrho_{1} > \varrho_{2}$ and
$\eta_{1} > \eta_{2}$, and examine flow in the northern hemisphere
$\theta < \pi/2$. Between the two fluids there exists a surface
tension $\sigma$.  The flows are assumed to be irrotational, except at
the interface. Fluid 1 is injected into fluid 2 through an inlet
located at the sphere's north pole, at a given flow rate $Q$, which is
the area covered per unit time. Fluid 2 is simultaneously withdrawn,
at the same rate, through an outlet placed at the south pole.  The
acceleration of gravity is constant, represented by {\bf g}, and
points from north to south pole.

During the flow, the fluid-fluid interface has a perturbed shape
described as $\rho={\cal R} \equiv R(t) + \zeta(\varphi,t)$. The
interface perturbation amplitude is represented by $\zeta(\varphi,t)$,
and $R$ denotes the time-dependent unperturbed radius 
\begin{equation}
\label{R}
R(t)=a \arccos \left ( C_{0} - \frac{Qt}{2 \pi a^{2}} \right ),
\end{equation}
where $C_{0}=\cos (R_{0}/a)$, and $R_{0}$ is the unperturbed radius at
$t=0$. The unperturbed shape is a polar cap of geodesic radius
$\rho=R$, surface area ${\cal A}=4\pi a^{2}
\sin^{2} (R/2a)$ and circumference ${\cal L}=2\pi a\sin{(R/a)}$. 

We express the net perturbation $\zeta(\varphi,t)$ as
a Fourier series
\begin{equation}
\label{z}
\zeta(\varphi,t)=\sum_{n=-\infty}^{+\infty} \zeta_{n}(t) \exp{(i n \varphi)}, 
\end{equation}
where $\zeta_{n}(t)$ denotes the complex Fourier mode amplitudes, and
$n$=0, $\pm 1$, $\pm 2$, $...$ is the discrete azimuthal wave number.
The area of the perturbed shape is kept independent of the
perturbation $\zeta$ by expressing the zeroth Fourier mode
in~(\ref{z}) as $\label{z0}
\zeta_{0}(t)= (-1/2a) ~\cot{(R/a)} ~\sum\limits_{n \neq 0} |\zeta_{n}(t)|^{2}$.

We consider a generalized version of Darcy's law~\cite{Saf,Rev},
adjusted to describe flow between concentric spheres
\begin{equation}
\label{Darcy}
{\bf v}_{j}= - \frac{b^{2}}{12\eta_{j}} \left [ { \nabla} p_{j} - \varrho_{j} ~g \sin \left ( \frac{\rho}{a} \right ) \hat{\rho} \right ],
\end{equation}
where ${\bf v}_{j}={\bf v}_{j}(\rho,\varphi)$ and
$p_{j}=p_{j}(\rho,\varphi)$ are, respectively, the velocity and
pressure in fluids $j=1$ and $2$. The gradient in
equation~(\ref{Darcy}) is associated with the metric~(\ref{metric})
and is obtained from the corresponding three-dimensional expression
for the gradient in spherical coordinates $(r,\theta,\varphi)$, by
keeping $r=a$ and noting that $\theta=\rho/a$. In contrast to
gravity-driven flows in flat Hele-Shaw cells, the gravity term
in~(\ref{Darcy}) is not constant, but depends on the radial 
distance $\rho$. This is a manifestation of the cell 
spatial curvature together with its embedding in three dimensional space.

We can exploit the irrotational flow condition to define the velocity
potential ${\bf v}_{j}=-{\bf \nabla} \phi_{j}$.  Using the velocity
potential, we evaluate equation~(\ref{Darcy}) for each of the fluids
on the interface, subtract the resulting equations from each other,
and divide by the sum of the two fluids' viscosities to get
\begin{equation}
\label{dimensionless2}
A \left ( \frac{\phi_{1}|_{{\cal R}} + \phi_{2}|_{{\cal R}}}{2} \right ) -  \left ( \frac{\phi_{1}|_{{\cal R}} - \phi_{2}|_{{\cal R}}}{2} \right ) = - \alpha ~(\kappa)|_{{\cal R}} - \gamma a \cos \left (\frac{{\cal R}}{a} \right ),
\end{equation}
where A=$(\eta_{2} - \eta_{1})/(\eta_{2} + \eta_{1})$ is the viscosity
contrast, $\alpha=b^{2} \sigma/[12(\eta_{1} + \eta_{2})]$ contains the
surface tension, and $\gamma=b^{2} g ~(\varrho_{1} -
\varrho_{2})/[12(\eta_{1} + \eta_{2})]$ is a measure of gravitational
force. To obtain~(\ref{dimensionless2}) we used the pressure boundary
condition $p_{2} - p_{1}=\sigma\kappa$ at the interface $\rho={\cal
R}$, where $\kappa$ is the interfacial curvature~\cite{Parisio}.

Following steps similar to those performed in Ref.~\cite{Parisio}, we
define Fourier expansions for the velocity potentials, which obey
Laplace's equation. We express $\phi_{j}$ in terms of the perturbation
amplitudes $\zeta_n$ by considering the kinematic boundary condition
for flow on a sphere. As in the flat cell case, this condition refers
to the continuity of the normal velocity across the fluid-fluid
interface~\cite{Ros}. Substituting these relations into
equation~(\ref{dimensionless2}), and Fourier transforming, yields the
mode coupling equation of the Saffman-Taylor problem in a spherical
Hele-Shaw cell, taking into account both injection and gravity
\begin{equation}
\label{result}
\dot{\zeta}_{n}=\lambda(n) ~\zeta_{n} + 
\sum_{n' \neq 0} \left [ F(n, n') ~\zeta_{n'} \zeta_{n - n'} + 
G(n, n') ~\dot{\zeta}_{n'} \zeta_{n - n'} \right ],
\end{equation}
where
\begin{equation}
\label{growth}
\lambda(n)= \left [ \frac{Q}{2 \pi a^{2}S^{2}} 
\left (A |n| - C \right ) - \frac{\alpha}{a^{3}S^{3}} |n| (n^{2} - 1) + 
|n|~\frac{\gamma}{a}\right ]
\end{equation}
is the linear growth rate, and
\begin{equation}
\label{F}
F(n, n')=\frac{|n|}{aS} \left \{ \frac{QAC}{2 \pi a^{2}S^{2}} 
\left [ \frac{1}{2} - sgn(nn') \right ] - \frac{\alpha C}{a^{3}S^{3}} 
\left [ 1 - \frac{n'}{2} ( 3 n' + n )  \right ] + 
\frac{\gamma C}{2a} \right \} + \frac{Q}{4 \pi a^{3} S},
\end{equation}
\begin{equation}
\label{G}
G(n, n')=\frac{1}{aS} \left \{ A|n| [ 1 - sgn(nn') ] - C \right \}
\end{equation}
are the second-order mode coupling terms, with $S=\sin(R/a)$ and
$C=\cos(R/a)$. The overdot denotes total time derivative and 
the sign function $sgn(nn')=1$ if $(nn') > 0$ and $sgn(nn')=-1$ 
if $(nn') < 0$. 
\section{Discussion}
\label{discussion}

\subsection{Linear stability analysis with $Q=0$}
\label{linear}

First concentrate on the purely gravity-driven case, $Q=0$. In this
situation, we consider flow in a closed cell obeying mass
conservation. To drive the interface gravitationally, one could first
allow the system to form a stable, unperturbed spherical cap at the
south pole, and then invert the spherical cell to put the denser fluid
on top in the unstable position at the north pole.

We begin by investigating the dispersion
relation~(\ref{growth}). Notice that the gravity term in
Eq.~(\ref{growth}) contains no explicit dependence on radial distance. This
finding is somewhat surprising, since the gravity term in Darcy's
law~(\ref{Darcy}) clearly presents a dependence on $\rho$ . Physical
intuition suggests a factor of $S$ should multiply the gravity term
in~(\ref{growth}), making it vanish at the poles and become maximal at
the equator.  This apparent missing factor reappears if we rewrite the
linear growth rate in terms of the wave number $k=|n|/aS$. The
variables $n$ and $k$ are both useful: for example, $n$ occurs in
integer values and counts the number of fingers. On the other hand,
$k$ determines the characteristic wavelength of a perturbation.

To better understand the physical information behind the description
of the linear stage in terms of $n$ and $k$, we plot the linear growth
rate at a sequence of radial distances $R=a \theta$ in figures 2a and 2b. We use
typical experimental parameters given in a recent experimental work in
rotating, flat Hele-Shaw cells~\cite{Carr}: fluid 1 is a silicone oil
($\eta_{1} \approx 0.5$ $\rm{g}$/$\rm{cm}$ $\rm{s}$, $\varrho_{1}
\approx 1.0$ $\rm{g}$/$\rm{cm^{3}}$) and fluid 2 is 
air ($\eta_{2} \approx 0$, $\varrho_{2} \approx 0$).  The thickness of
the cell $b=0.1$ $\rm{cm}$ and the surface tension $\sigma=20.7$
$\rm{dyne/cm}$. We set the radius of the sphere $a=5$ $\rm{cm}$ and
acceleration of gravity $g=980$ $\rm{cm}$/$\rm{s^{2}}$.

Comparing figures 2a and 2b, it is clear that the fastest growing mode
(number of fingers $n$ developing at a maximum growth rate) moves to
large $n$ for large radial distance (and small $n$ for small radial
distance) in order to keep the corresponding fastest growing $k$, and
thus the apparent wavelength, nearly constant.  The same effect occurs
in radial flow in flat space~\cite{MW}.  In addition to the shift to
larger $n$, the growth rate of the fastest growing mode increases with
radial distance because the net gravitational force depends on $R$.
This differs from the case of radial flow in flat space~\cite{MW},
where the interface velocity falls off as $1/R$ causing the growth
rate of the fastest growing mode to decrease.

Consider the purely linear contribution, which appears as the first
term on the right hand side of equation~(\ref{result}).  The condition
$Q=0$ simplifies the theoretical description: $R(t)=R_{0}=$ constant,
and consequently the linear growth rate $\lambda(n)$ is {\it time
independent}. This implies that the actual relaxation or growth of
mode $n$ is purely exponential
\begin{equation}
\label{relax}
\zeta_{n}^{(Q=0)}~(t)=\zeta_{n}(0) ~\exp[\lambda(n)~t],
\end{equation}
where $\zeta_{n}(0)$ is the initial perturbation amplitude. 
To see the overall effect of Eq.~(\ref{relax}), we plot evolved
interfaces using the same experimental parameters 
as those used in figures 2a and 2b. It is convenient to rewrite 
the net perturbation~(\ref{z}) in terms of cosine and sine modes 
$\zeta(\theta,t)= \zeta_{0} + \sum\limits_{n > 0} 
\left[ a_{n}(t)\cos(n\theta) + b_{n}(t)\sin(n\theta) \right ]$,
where $a_{n}=\zeta_{n} + \zeta_{-n}$ and $b_{n}=i \left ( \zeta_{n} -
\zeta_{-n} \right )$ are real-valued. We take into account modes 
$n$ ranging from $n=1$ up to $20$. Figure 3 depicts the evolution of
the interfaces, for a random choice of phases, at time $t=5$
$\rm{s}$. We evolve from two distinct initial radii (a) $R_{0}=a
~\pi/8$, and (b) $R_{0}=a ~\pi/4$. In both cases $|\zeta_{n}(0)|=0.05$
$\rm{cm}$ and we use the same randomly chosen phases. It is evident
that for larger radial distances (or equivalently, larger polar angles) 
we have both faster growth and more unstable
modes, though the characteristic wavelength is not very different, in
agreement with the predictions made from figures 2a and 2b.

\subsection{Linear stability analysis with $Q>0$}
\label{inject}

Now consider the nonvanishing injection case. Taking $Q>0$ introduces
two important new effects. First, inspecting equation~(\ref{growth}),
we see that $Q$ multiplies a term linear in $n$ that must be added to
the growth rate $\lambda$ obtained with $Q=0$. If the inner fluid
is high viscosity, so that $A<0$, then this new term in $\lambda$ is
negative. All modes grow more slowly with $Q>0$, and some may even
become stable. Thus, $Q>0$ diminishes the strength of the instability.
Essentially, the flow fills in the gaps between the fingers.

The second effect is more subtle. Since the unperturbed interface
radius $R(t)$ is now {\it time dependent}, the linear growth rate
$\lambda$ evolves with time as the radial distance steadily grows.  At
any instant the interface evolves exponentially with an instantaneous
growth rate $\lambda$ depending on the current $R(t)$. For
sufficiently small $Q$ the instantaneous $\lambda$ at $R(t)$ nearly
equals the $Q=0$ value for the same $R(t)$. We call this a {\em
quasistatic} approximation, and within this approximation it is clear
that an increasing number of modes become unstable as time progresses
due to the steady increase of $R(t)$.  The evolution of $\lambda$ in
the quasistatic approximation is given by the series of curves shown
in figure 2.

If we wish to study the linear growth {\em without} making the
quasistatic approximation, we may integrate the equation of
motion~(\ref{result}) exactly, keeping only terms of first order in
$\zeta$ on the right-hand side.  Because $\lambda$ is time dependent,
through the variation of $R(t)$, the exact solution of the linear
equation of motion becomes non-exponential and can be written as
\begin{equation}
\label{linearsol}
\zeta_{n}^{(Q > 0)}~(t)=\zeta_{n}(0) ~\frac{S_{0}}{S} \left [\frac{\tan{(R/2a)}}{{\tan(R_{0}/2a)}} \right ]^{A|n|} \exp \left \{ 
\frac{2 \pi |n|}{Q} \left [ \frac{\alpha (n^{2} - 1)}{a} \left ( \frac{C}{S} - \frac{C_{0}}{S_{0}} \right ) + \gamma a \left ( C_{0} -  C \right ) \right ] \right \} \end{equation}
where $S_{0}=\sin{(R_{0}/a)}$, and $R$, $C$ and $S$ all depend on
time.

\subsection{Mode coupling analysis}
\label{nonlinear}

We use the mode coupling equation~(\ref{result}) to investigate 
the effect of the nonlinear terms in the interface evolution. 
We are interested in studying how gravity influences the shape 
of the fingering structures, focusing on finger-tip behavior. 
Finger tip-splitting and tip-sharpening phenomena are related to 
the influence of a fundamental mode $n$ on the growth of its harmonic $2n$. 
Without loss of generality we may choose the phase 
of the fundamental mode so that $a_{n} > 0$ and
$b_{n}=0$.  We replace the time derivative terms $\dot{a}_{n}$ and
$\dot{b}_{n}$ by $\lambda(n)~a_{n}$ and $\lambda(n)~b_{n}$,
respectively, for consistent second order expressions. Under these
circumstances the equation of motion for the harmonic 
cosine mode becomes
\begin{equation}
\label{new3}
\dot{a}_{2n}=\lambda(2n)~a_{2n} + \frac{1}{2} ~T(2n,n) ~a_{n}^2
\end{equation}
where the finger-tip function is 
\begin{equation}
\label{T(2n,n)}
T(2n,n)=\frac{2n}{aS}~\left \{ \frac{Q}{2 \pi a^{2}S^{2}} \left [ \frac{(C^{2} + 1)}{4n} - AC \right ] + \frac{3 \alpha C ~(2n^{2} - 1)}{2a^{3}S^{3}} \right \}.
\end{equation}
The corresponding equation for the sine modes $b_{2n}$ is not 
as interesting as~(\ref{new3}), since growth of $b_{2n}$ is 
uninfluenced by $a_{n}$. 

The sign of $T(2n,n)$ dictates whether finger tip-splitting or finger
tip-sharpening is favored by the dynamics~\cite{MW}. If $T(2n,n)<0$,
at second order the result is a driving term of order $a_{n}^{2}$
forcing growth of $a_{2n} < 0$.  With this particular phase of the
harmonic forced by the dynamics, the $n$ outwards-pointing fingers of
the fundamental mode $n$ tend to split. In contrast, if $T(2n,n)>0$
growth of $a_{2n} > 0$ would be favored, leading to outwards-pointing
finger tip-sharpening.

A noteworthy point about equation~(\ref{T(2n,n)}) is that it shows no
dependence whatsoever on gravity. In the evaluation of $T(2n,n)$ from
equation~(\ref{result}) we found that the term involving gravity in
$F(2n,n)$ exactly cancels against the term involving gravity in
$\lambda(n)G(2n,n)$. The second order term driving tip-splitting in
equation~(\ref{new3}) is therefore independent of the force of
gravity, though gravity does generally influence mode coupling
at second and higher orders.

Inspecting Eq.~(\ref{T(2n,n)}) we find that, since $A=-1$ and $C>0$,
the finger-tip function $T(2n,n)>0$ for both $Q=0$ and
$Q>0$. Eq.~(\ref{T(2n,n)}) predicts that gravity-driven flow on a
sphere leads to patterns showing enhanced finger tip-narrowing.
Informal studies of chocolate syrup fingers on ice cream scoops
support this claim.

\vspace{0.5 cm}
\begin{center}
{\bf ACKNOWLEDGMENTS}
\end {center}
\noindent
J.A.M., F.P., and F.M. thank CNPq and FINEP for financial support. 
Work of M.W. was supported in part by the National Science Foundation 
grant No. DMR-9732567.

\pagebreak
\noindent
\centerline{{\large {FIGURE CAPTIONS}}}
\vskip 0.5 in
\noindent
{FIG. 1:} Schematic configuration of flow in a spherical Hele-Shaw cell. 
\vskip 0.5 in
\noindent
{FIG. 2:} Linear growth rate as a function of (a) mode number $n$ and 
(b) wave number $k$ at a sequence of radial distances (1) $R=a ~\pi/8$, (2) $ R=a ~\pi/4$, (3) $R=a~3\pi/8$ and (4) $R=a~\pi/2$.
\vskip 0.5 in
\noindent
{FIG. 3:} Interface evolution according to Eq.~(\ref{relax}) 
at $t=5$ $\rm{s}$, for (a) $R_0=a~\pi/8$ and (b) $R_0=a~\pi/4$. 
Other parameters are given in the text. 

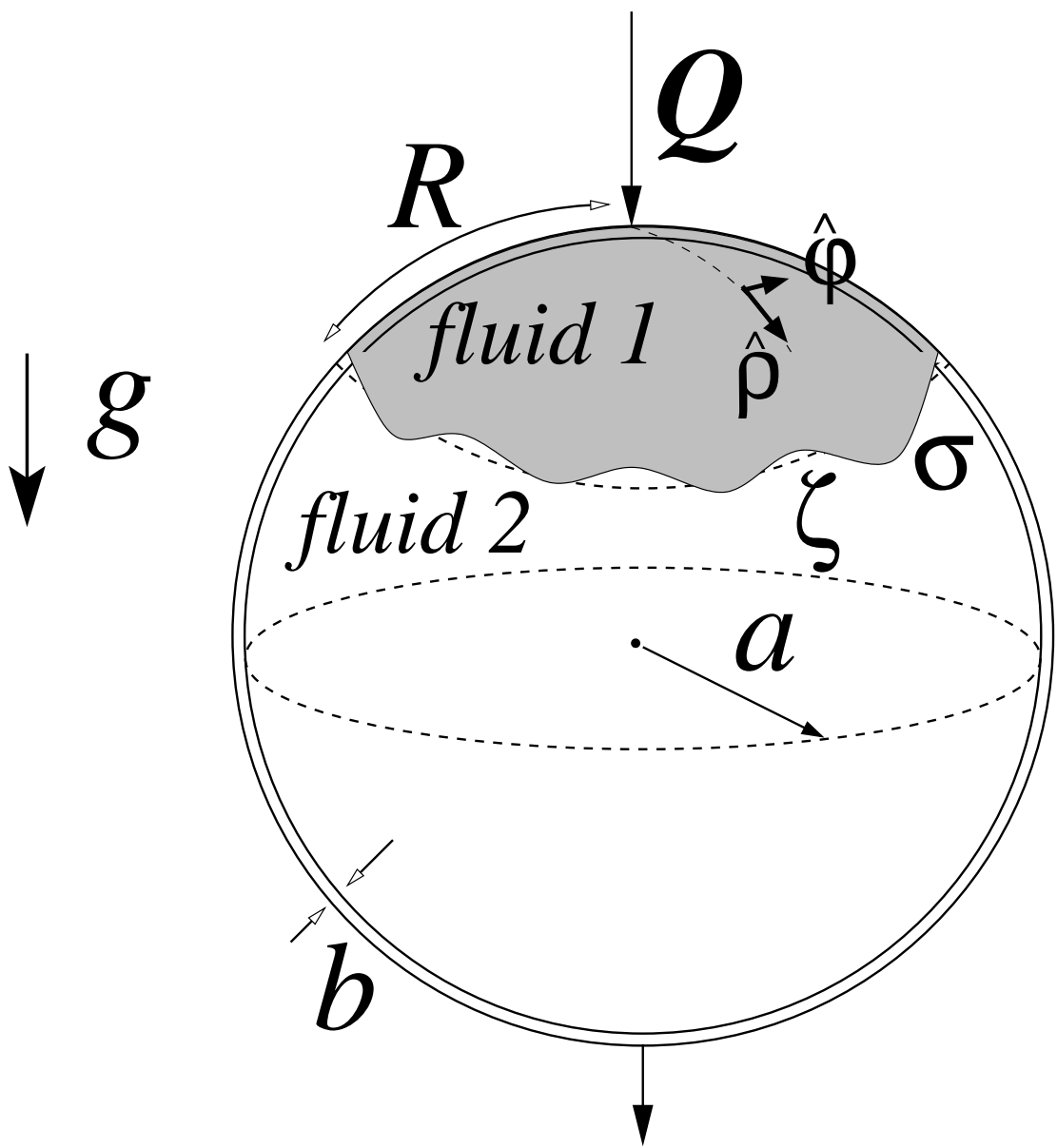
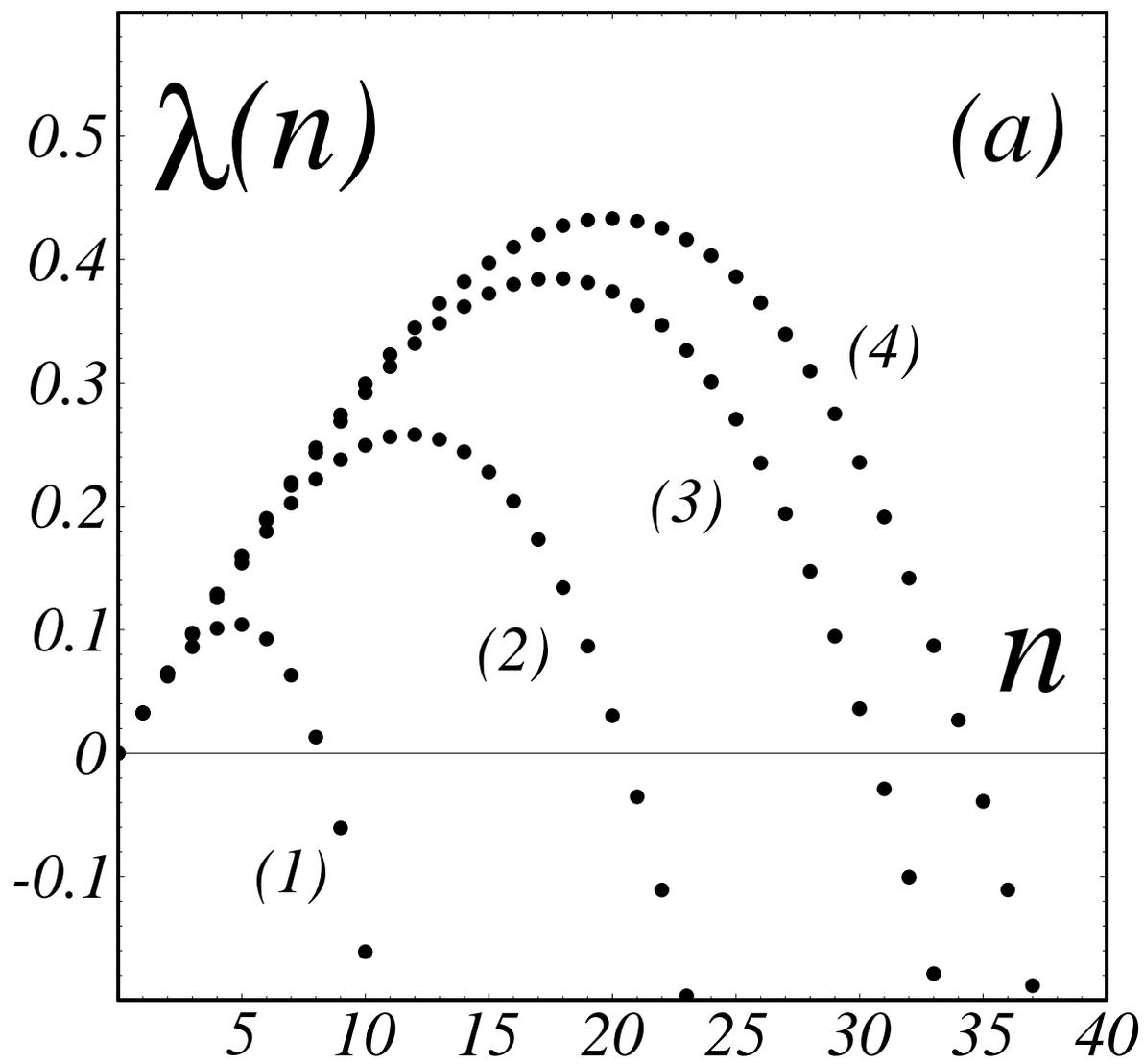
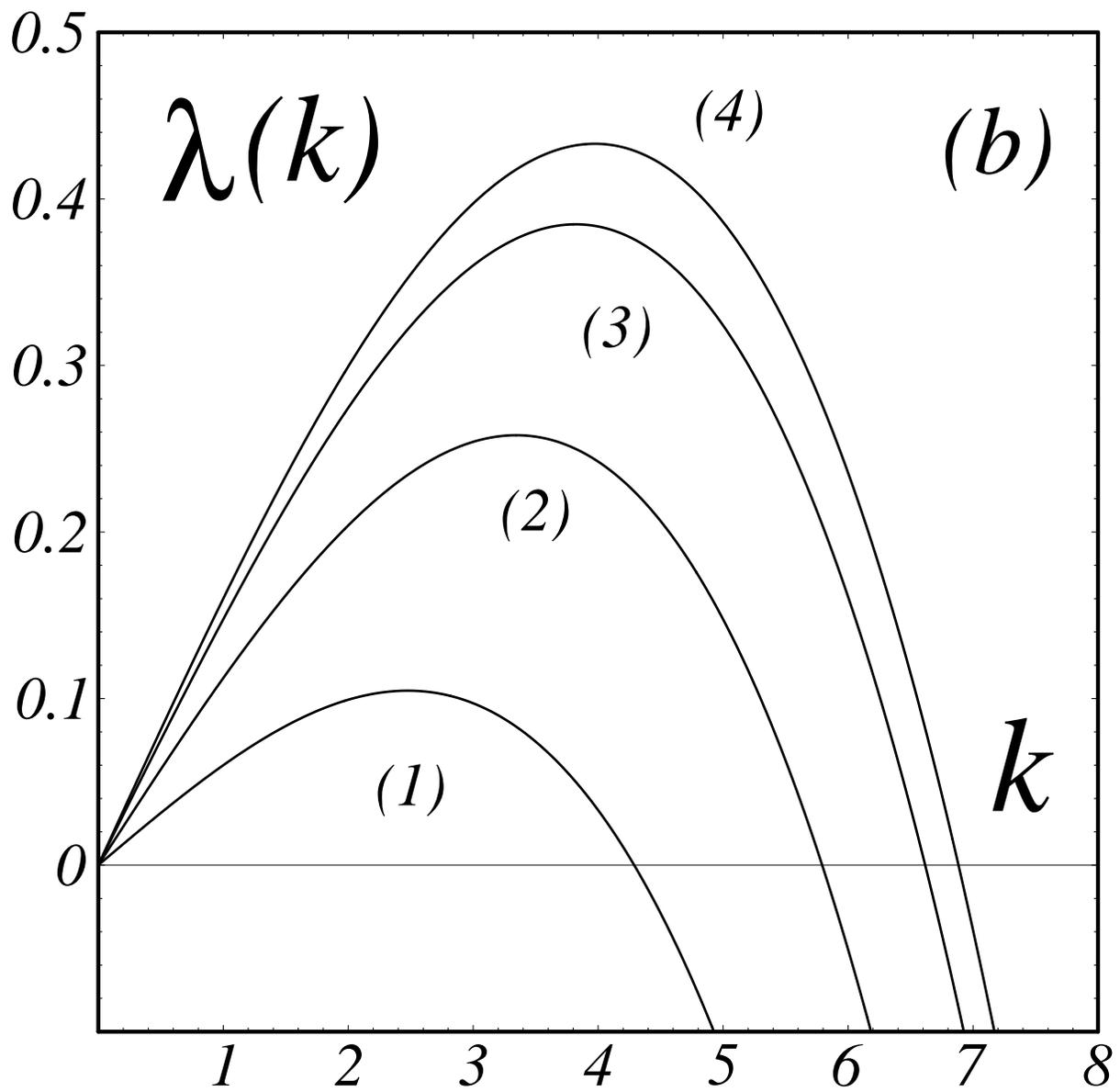
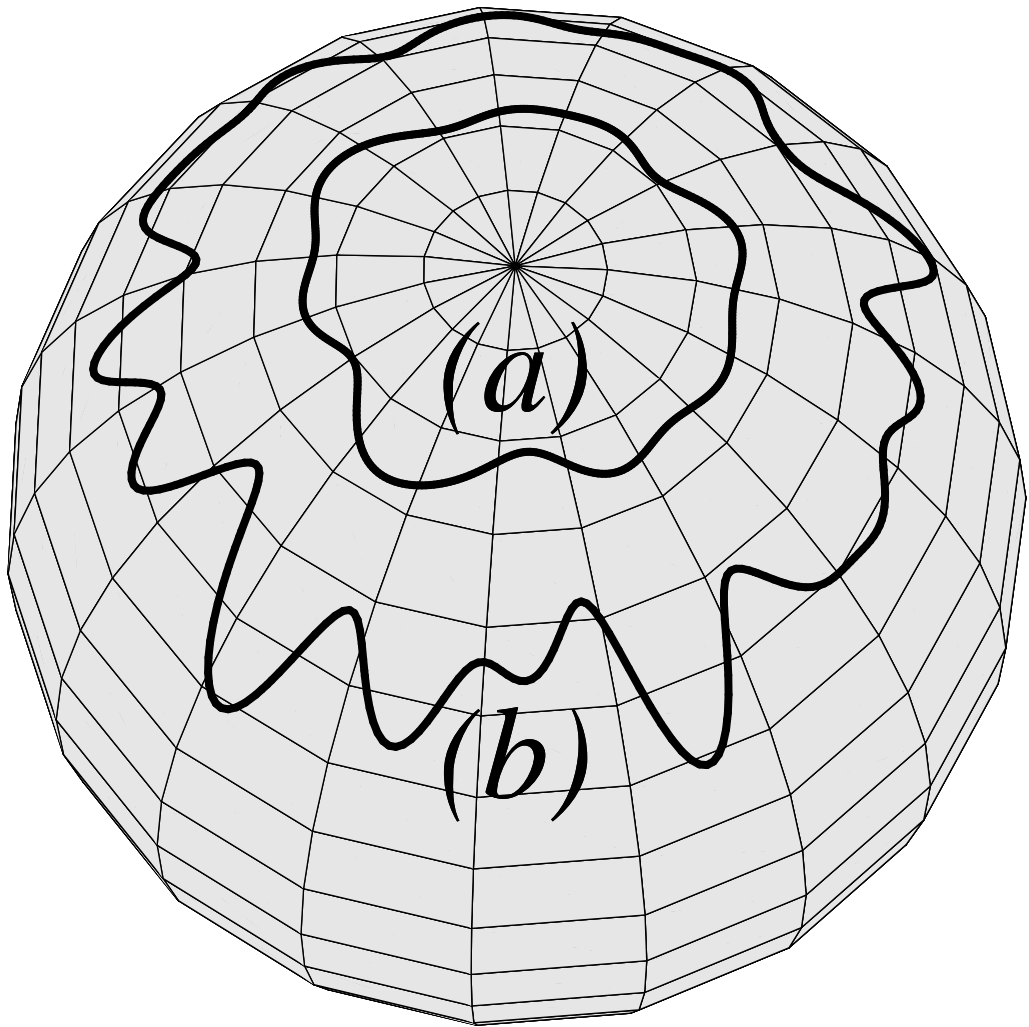
\end{document}